\documentclass[%
 aip,
 amsmath,amssymb,
 reprint,%
]{revtex4-1}

\usepackage{graphicx}
\usepackage{dcolumn}
\usepackage{bm}

\usepackage[utf8]{inputenc}
\usepackage[T1]{fontenc}
\usepackage{mathptmx}

\begin{document}

\preprint{AIP/123-QED}

\title[Effects of Electromagnetic Acceleration on the StarOne C1 Satellite Using Planar Antenna Array]{Effects of Electromagnetic Acceleration on the StarOne C1 Satellite Using Planar Antenna Array}

\author{A. Heilmann}
  \email{heilmann@ufpr.br}
\author{H. Tertuliano Filho}%
 \email{tertulia@eletrica.ufpr.br}

\author{C.A. Dartora}
 \email{cadartora@eletrica.ufpr.br}

\author{D. Cl\'istenes}%
 \email{david.clistenes@ufpr.br}

\author{A. M. Adams}%
 \email{augusto.adams@ufpr.br}
\affiliation{ 
Department Electrical Engineering - Group of Signal Propagation System (SPS) and Atmospheric Electricity Phenomena (FEA), Federal University of Paran\'a, C.P. 19011, 81531-990, Curitiba-PR, Brazi
}%

\date{\today}

\begin{abstract}
Artificial satellites perform down link communication or transmission with earth stations. These transmissions are made from an array of antennas chosen to increase directivity and gain in the transmission of the electromagnetic signal. It occurs that the signal from the array of antennas can infer a disturbance by electromagnetic acceleration, which results in disturbances in the orbital components of the satellite. Taking the StarOne C1 satellite, with EIRP = 44 W, mass = 1918 kg  and with a planar antenna array, we developed an acceleration model for this type of antennas. A satellite orbiter propagator was developed by the Runge-Kuta method, and from the technical characteristics of this satellite and the state vector values (position and velocity), we were able to calculate an electromagnetic acceleration of the order of $10^{-9}$ m$/$s$^{2}$. 
\end{abstract}

\maketitle

\section{\label{sec:level1}Introduction}
Artificial satellites have well-defined orbits around the Earth. According to a more recent definition, the trajectory of a satellite can be low orbit 80 km up to 2.000 km from the Earth's surface, medium orbit between 2.000 $-$ 35.786 km or high earth orbit, over 35.786 km. For each orbit, the satellite supports specific functions and characteristics such as telecommunications (LEO), Monitoring, Global Positioning (MEO) and Geostationary and Remote Sensing (HEO) \cite{Mon}. 

Satellite parameters can be seen as their state vectors can be perturbed by Moon and Sun (in the order of 5 $\times$ 10$^{-6}$ m/s$^{2}$), due to tidal forces (in order of 1 $\times$ 10$^{-9}$ m/s$^{2}$), due to direct solar radiation pressure (in the order of 1 $\times$ 10$^{-7}$ m/s$^{2}$), and due to atmospheric drag (less of 1 $\times$ 10$^{-9}$ m/s$^{2}$) \cite{Mon}, \cite{Peter}.

The effects this perturbations can be an increase of the semi-axis major, of the eccentricity, of the period, and changes in the other orbital elements an so, increasing inclination and the perigee argument \cite{Mon}. Disturbances of electromagnetic origin were also presented magnitude order for acceleration over a satellite \cite{Armando}.


Considering that the down link communication between the satellites and an antenna on the terrestrial surface is carried out from an antennas array, an electromagnetic disturbance model was implemented which considers a planar antenna array over a satellite with characteristics of the StarOne C1 satellite, with the mass of 1918 kg and EIRP of 44 W. \cite{Allnutt}, \cite{Ainoa}. 

\section{Antenna Arrays for Satellite Communication}

For antenna array, the beam width in the plane perpendicular to the axis of the set is determined by the beam width of the elements of this plane. Advances in construction and integrated electronics for power supply, the cost for the development of antennas in the form of array has become feasible, especially for communications with satellites.

Main advantages of antenna arrays over large reflectors are the higher flexibility, lower
production and maintenance cost, modulate and a more efficient use of the spectrum \cite{Iida}. The radiation efficiency of an array antenna is normally much lower than that of a reflector antenna of comparable dimensions.



Considering an flat surface on which the antenna elements, the square perimeter, uniform spacing between lines and columns of your elements, is possible to use the principle of multiplication of the diagram for co linear set, in which the diagram of electric field of the similar set is the product of diagram of the isolated element by punctual isotropic amplitudes and phases relative to the original set (figure (1)).


\begin{figure} [!h]
	\begin{center}
		\scalebox{0.33}{\includegraphics{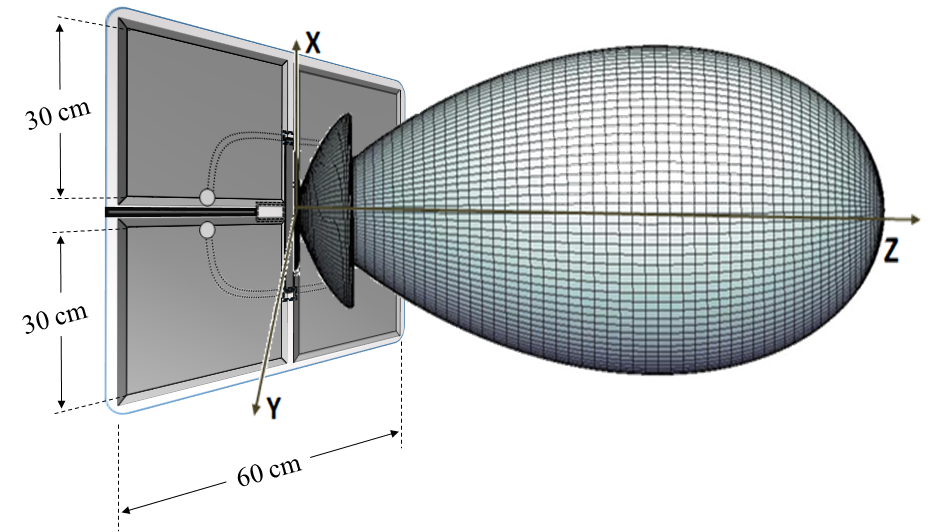}}
	\end{center}
	\caption{Radiation diagram of a planar antenna of 4 elements with dimensions. The z-axis points toward the Earth surface.}
	\label{fig1}
\end{figure}

%
%
%

Planar arrays are used in antennas embedded in satellites by allowing a highly directional beam. Considering a cell of an array, which is the area surrounding an element, it is possible to derive from the antenna theory the directivity of an ideal dipole with $N$ elements of area $A_{Cel}$ in which:
\begin{equation}
D_{MaxCel} = N \frac{4 \pi}{\lambda^{2}} A_{Cel}.
\end{equation}

This equation indicates that an array of antennas in a plane $x-y$ has directivity proportional to its area, $A_{Cel}$. As a function of the angle, the directive varies with the projection of the area cos ($\theta$), that to seen from an angle $\theta$ in relation to the direction of broadside, that is, planar array broadside accompany a maxima irradiation in the normal direction of the axis of the arrangement \cite{Planar}.

Since the Poynting vector modulus is the ratio of the squared module of the electric field vector to the impedance in vacuum ($Z_{o} = 377 \Omega$), so:
\begin{equation}
\vec{S}_{rad}(r, \theta, \phi) = \frac{|\vec{E}_{rad}|^{2}}{2Z_{o}},
\end{equation}


In antenna theory one takes maximum angular directivity as the definition for directivity:
\begin{equation}
D = max [F(\theta, \phi)].
\end{equation}
The radiated total power ($ P_{rad}$) by the antenna array is defined as $ P_{rad} = \int_{\Omega} r^{2} S_{rad} d \Omega$, where $d \Omega$ is the infinitesimal element of the solid angle. Therefore, the gain of an array of antennas is defined as a measure of the concentration of the effective power isotropically radiated in a given direction ($ \theta, \phi $).

\section{Choice of Satellite Communication}

StarOne C1 marked the beginning of operations of Embratel StarOne third generation of satellites (C Series). The new satellite replaced Brasilsat B2 in the 65.0$^{o}$ W orbital position and has almost twice as much power than its predecessor. StarOne C1 satellite covers the entire territory of Brazil and the South and Central American countries as well, in addition to Florida, US. 




During the normal lifetime the C1 satellite will be maintained in orbit with a tolerance of $\pm$ 0.07$^{o}$ in latitude and longitude about the sub-satellite point. (The sub-satellite point is the point on the earth's surface directly below the nominal satellite position.) Earth station antennas of up to about 7 m in diameter accessing the Optus Ku-band frequencies of 14/12 GHz generally do not require tracking capability \cite{Allnutt}, \cite{Iida}, \cite{Ainoa}.  

%


The table I shows the technical characteristics of the StarOne C1 satellite.

\begin{table}
\caption{\label{tab:table1}Satellite technical data of StarOne C1.}
\begin{ruledtabular}
\begin{tabular}{lcr}
Description&Characteristic\\
\hline
Name & StarOne C1 \\
Type/Application & Communication \\
Mass & 1918 kg \\
Orbit & GEO \\
Vector ray & $~$35880 km \\
P{tx} & 40 dBW \\
Polarization & Vertical \\
Antenna & Array Planar \\
Elements Number & 4 \\
Frequency (Band C) & (5850 MHz up link-4200 MHz down link)\\
Frequency (Band ku) & (13750 MHz up link-12200 MHz down link)\\
\end{tabular}
\end{ruledtabular}
\end{table}

\section{Electromagnetic Acceleration Model for Planar Antenna Array}

From the antenna theory seen above, it is possible to represent the power density radiated by an antenna such as:
\begin{equation}
S_{rad} = \frac{G(\theta,\phi) P_{rad}}{4 \pi r^{2}},
\end{equation}
\noindent
where $G(\theta,\phi)$ is the ideal element gain, described in equation (5). 
considering yet the linear momentum conservation electromagnetic, as:
\begin{equation}
d \vec{p}_{em} = \frac{P_{rad}}{4 \pi c^{2}} dt \int_{\Omega} G(\theta, \phi) \hat{a}_{r} d\Omega.
\end{equation}

And taking into account that, $\vec{p}_{mec} = m\vec{v}$, and taking the momentum conservation, so,
\begin{equation}
\begin{aligned}
\frac{\vec{p}_{em}}{\Delta t} = \int_{V} \frac{G(\theta, \phi) P_{rad}}{4 \pi c} \hat{a}_{r} \sin\theta dt d\theta d\phi~.
\end{aligned}
\end{equation}

From this, we conclude that the expression on the left is by definition equal to the electromagnetic force, 
\begin{equation}
\vec{F}_{sat} = - \frac{d\vec{p}_{mec}}{dt},
\end{equation}
\noindent
that using Newton's Law we find:
\begin{widetext}
\begin{equation}
\vec{a}_{sat} = - \frac{P_{rad}G(\theta, \phi)}{4 \pi m_{sat} c} \int_{\theta =0}^{\theta _0}
\int _{\varphi =0}^{2 \pi} [\sin\theta \cos\varphi {\bf\hat {a}_x} + 
\sin\theta \sin \phi {\bf\hat {a}_y} + \cos\theta {\bf\hat {a}_z}]  \sin\theta d\theta d\phi~.
\end{equation}
\end{widetext}

By definition EIRP $=$ P$_{rad}$ G, then we consider that the total ideal element gain is normalized by a factor $ G $ such that $ G(\theta, \phi) = G(\theta, \phi) G  $ \cite{Notaros}. 

We integrate the coordinates of the solid angle in the direction $ \hat{z} $. Thus we can rewrite the equation of the ideal element gain as, $G(\theta, \phi) = \frac{4 \pi}{\lambda^{2}} A_{Cel} cos(\theta)$,

%
which including in expression of acceleration, results in, 
\begin{equation}
\begin{aligned}
\vec{a}_{sat} = - \frac{EIRP}{2  m_{sat} c} \int_{0}^{\pi} \frac{4 \pi A_{Cel}}{\lambda^{2}} sin(\theta)cos^{2}(\theta)d\theta.
\end{aligned}
\end{equation}

We solve the integral for $d\theta$ where, $\int_{0}^{\pi} sin(\theta)cos^{2}(\theta) d\theta= 2/3$, we have an expression for acceleration as: 

\begin{equation}
\vec{a}_{sat} \cong - \frac{47.039 EIRP}{m_{sat} c} ,
\end{equation}
\noindent
the constant 47.039 has dimension of [kg m/s]$^{-1}$ and the velocity of light on vacuum is c $=$ 3 $\times$ 10$^{8}$ m/s. So the electromagnetic disturbance model for an antenna array planar [m/s$^{2}$] is only function of the Equivalent Isotropically Radiated Power and satellite mass \cite{Notaros}, \cite{Armando}.

\section{Iterative System of Orbital Parameters}
To solve problems of ordinary differential equations, we will use the numerical algorithm$^{\textregistered}$ using the Runge-Kutta method, called ODE45 (Ordinary Differential Equation), propagated to a period of 48 hours, with step every 5 minutes of integration and absolute and relative tolerance of error, with 10$^{-14}$ of precision in the mantissa for position and velocity \cite{Armando}.

We analyze the components of the electromagnetic acceleration from the satellite antenna reference in the Normal (N), Transverse (T) and Radial (R) directions to the orbit plane, respectively. The force of the antenna is in the radial direction, so it occurs only for the radial component. From this, the radial component of the array of antennas will be given by:
\[R = \vec{a}_{sat}. \hat{a}_{z} = a_{sat}\begin{pmatrix} 0 \\  0 \\Z/r\end{pmatrix} \]
\noindent
where $ \vec {a}_{sat}$ represents the perturbation model from the satellite antenna, in the case of an array of planar antennas (eq.19), $ r = (X^{2} + Y^{2} + Z^{2})^ {1/2}$, with respect to the Z-axis, of the antenna, respectively due to the antenna radiation reaction.


The equation of motion of a satellite as a function of the disturbing force due to antenna radiation is:

\begin{equation}
\ddot{\vec{r}} - -\frac{\mu}{r^{3}} \vec{r} + \vec{a}_{sat},
\end{equation}

From this equation we also remove the orbital velocity values of the satellite, and therefore, after propagation for a finite number of days, we have all the position values and orbital velocities of the StarOne C1 satellite, including the disturbance due to electromagnetic radiation ($ \vec{a}_{sat}$) of the array of planar antennas.

The equations are solved by the Runge-Kutta method, whose input values in the routine are the so-called State Vectors, which carry the position information ($ X, Y $ and $ Z $) and velocity ($V_{x}, V_{y}$ e $V_{z}$) of StarOne C1, for the following set of Keplerian Elements (table II):

\begin{table}
\caption{\label{tab:table1}Keplerian set of StarOne C1 satellite for the day, 19/09/2018 - 18:42:20 (UTC).}
\begin{ruledtabular}
\begin{tabular}{lcr}
Keplerian Elements&Parameters\\
\hline
Semi-major axis (m) & 4.215 $\times$ 10$^{7}$\\
		Eccentricity & 0.0002573\\
		Argument of Perigeo ($^{o}$) & 53.5871\\
		Ascension straight of ascending node ($^{o}$) & 98.2308\\
		inclination ($^{o}$) & 0.0289\\
		Mean Anomaly ($^{o}$) & 208.18\\
		Angle of terrestrial rotation ($^{o}$) & 65.0108\\
\end{tabular}
\end{ruledtabular}
\end{table}

\section{Results and Discussion}

First, the orbit was propagated without any perturbation, then the orbital propagation was recalculated considering the model of electromagnetic acceleration (Eq. 16).
The comparison of the results between these two orbital propagations are analyzed by the positional deviations in the components Radial ($\Delta R$), Normal ($\Delta N$) and Transversal ($\Delta T$).
\begin{figure} [!t]
	\begin{center}
		\scalebox{0.21}{\includegraphics{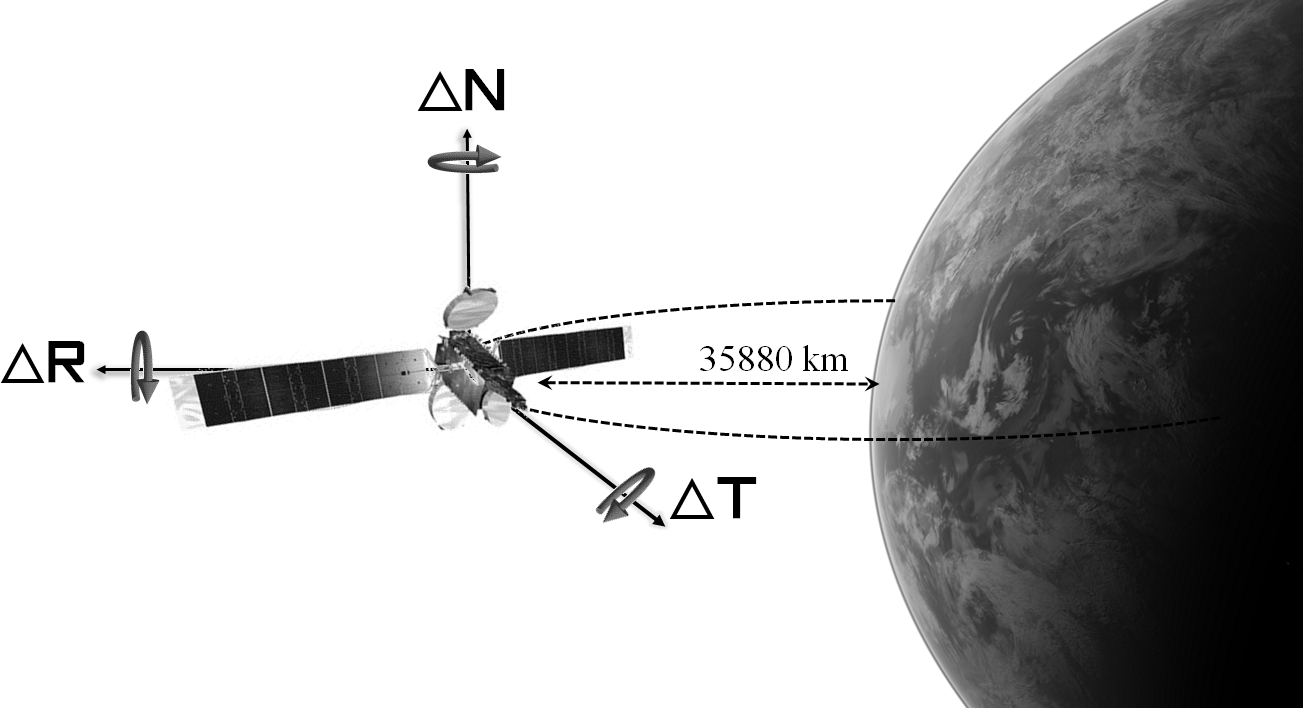}}
	\end{center}
	\caption{Deviations from orthogonal components: RADIAL, NORMAL e TRANSVERSAL.}
	\label{fig3}
\end{figure}

The normal component is perpendicular to the orbital plane and positive in the direction of the angular momentum vector. The radial orthogonal component is along the vector radius of the satellite and positive outside the central body, since the transverse component is in the direction of the orbital plane and positive in the direction of the satellite movement, as shown in figure (2).

As established by recommendation ITU-R P. 1546-4 the calculations for the EIRP are obtained through equation,
\begin{equation}
EIRP = P_{tx} + G_{tx} - \sum Loss_{tx},
\end{equation}
\noindent
where $G_{tx}$ the gain of the transmission antenna in relation to the array of antennas in dBd;
P$_{tx}$ nominal transmitting power specified by the manufacturer in dBW;
$\sum Loss$ is the sum of all losses in dB, found in transmission components, transmission lines, combiners and connectors. According to table I, P$_{tx} =$ 40 dBW, and making the gain equal to 12 dBd, the EIRP will be 44 W.

This parameter plus one mass (1918 kg) inserted in the propagator, we have the state vectors (position [m] and velocity [m/s]), from the numerical algorithm called $SAT-LAB$, graphical user interface for simulating and visualizing keplerian satellie orbits. 


Assuming an planar antennas array embedded in the satellite StarOne C1, a EIRP$=$44 Watt, the dimensions of the array of antennas according to figure(1),  $M_{sat} = 1918$ kg, total area of antenna array A$_{Cel}=$0.36 m$^{2}$, during a communication of \textit{down link} at band C, the frequency will be 4200 MHz, it was possible to find an acceleration of the order of,
\[|{\bf a}| \cong - \frac{47.039 EIRP}{m_{sat} c} = 3.6 \times 10^{-9} {\rm m/s}^2~. \]
We see that such effect is non negligible, being in the same order of magnitude of other perturbations previously cited.


The largest deviations were observed on the Radial, Normal and Transverse components (figure (3)) and on the X, Y and Z coordinates of the satellite figure (4).
\begin{figure} [!h]
	\begin{center}
		\scalebox{0.32}{\includegraphics{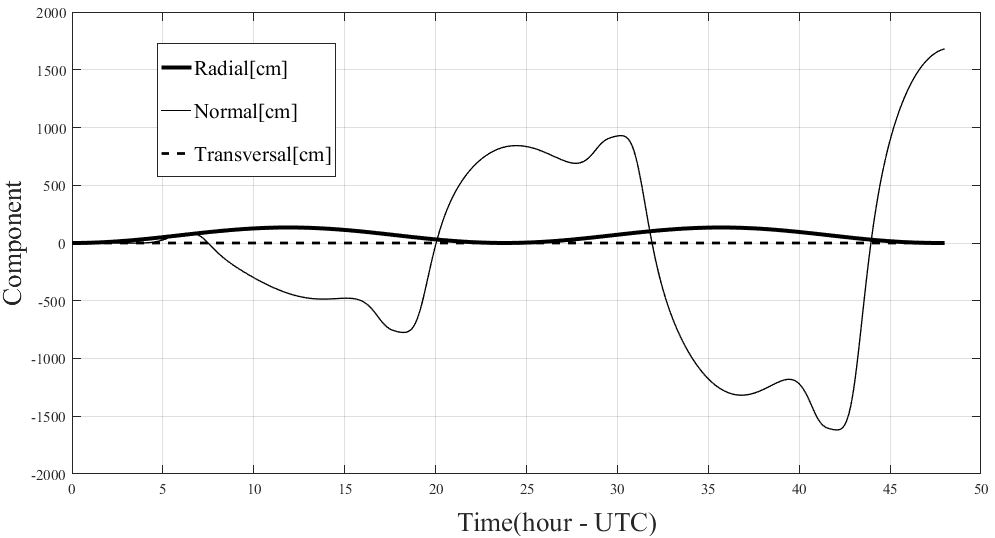}}
	\end{center}
	\caption{Deviation of at components Normal (thin line), Radial (thick line) and Transverse (dotted) of the StarOne C1 satellite.}
	\label{fi5}
\end{figure}

For a period of 48 hours, there was no significant variation in the transverse component, however, the radial component showed an oscillation every 12 hours, with return to the zero variation [cm], this justifies why the variation in altitude (vector radius) also is a sinusoidal, with maximum variation between 0 $-$ 130  km. The normal component presented a bi modal variation, with a maximum of $ \approx $ 1700 cm in 48 hours of propagation, which means that it can directly changes the angular momentum of the satellite. 

\begin{figure} [!h]
	\begin{center}
		\scalebox{0.275}{\includegraphics{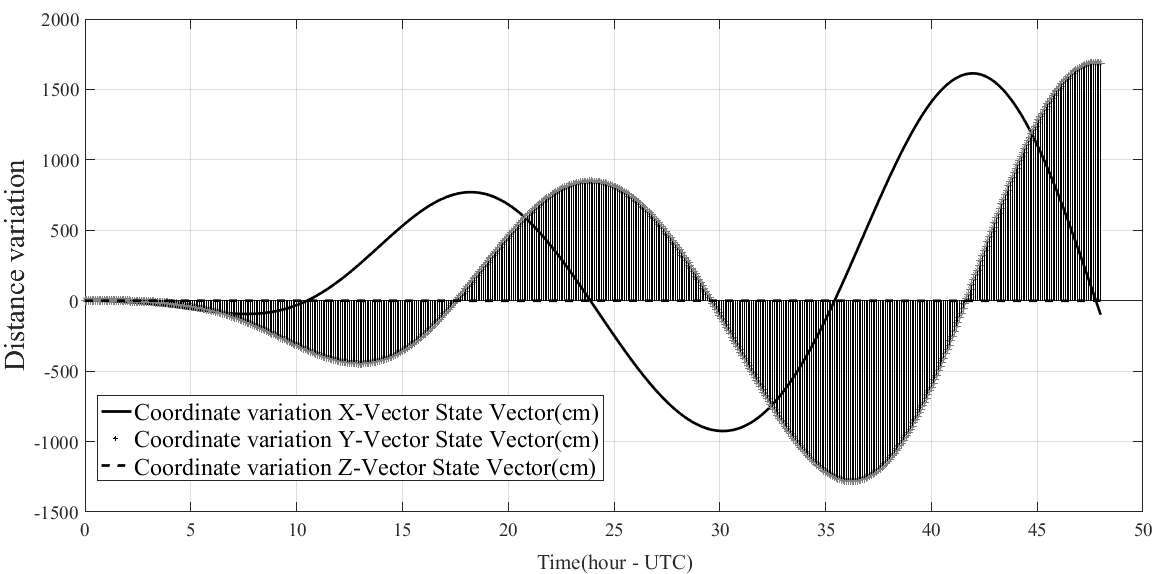}}
	\end{center}
	\caption{Variation in the X (thick line), Y (Stem) and Z (dotted) coordinates of the position vector of the StarOne C1 satellite.}
	\label{fi6}
\end{figure}

About the coordinates of the position state vector, there was no variation of the distance in cm in the Z coordinate, but in the Y (Stem) coordinate the deviation was $ \approx $ 1700 cm, followed by a variation in the coordinate X (thick line) that reached $ \approx $ 1600 cm.

We observed that the electromagnetic acceleration perturbation model imposes larger deviations for the normal (N), and coordinate (Y) components, with electromagnetic reaction acceleration of  $\approx$3.6 $\times$ 10$^{-9}$ m/s$^{2}$. For EIRP value 2 times greater, the variations in the normal component (N) and in the Y coordinate, increase of 300 cm, respectively. 

\section{Conclusion}

In summary, in this paper we developed a model to describe the effects of radiation reaction in satellite orbits due to the emitted RF power by the satellite transmitting antennas array planar. We demonstrated that the satellite acceleration can be in the range of  $\sim$ 10$^{-9}$ m/s$^2$, which is similar to other small order effects that must be taken into account in order to correct the satellite trajectory. 
The values of the deviations in the radial and transverse directions are irrelevant in propagation of the orbit of this satellite, this implies in the fact that there is also no deviation from the larger semi-axis, which depends on the radial component.
The slope of the orbit of the StarOne C1 satellite depends on the normal component, but in the simulations no variation of the inclination of the orbit was identified. This is because the deviations presented in the normal component (N) are symmetrical with a sinusoidal profile, which, although increasing over time, reestablishes any orthogonal slope discrepancies.
The largest variations occur on the coordinates of the position state vector, which can reach up to 1700 cm for the Y coordinate, after 2 days propagation, with electromagnetic reaction acceleration of $\approx$ 3.6 $\times$ 10$^{-9}$ m/s$^{2}$, being necessary an orbital correction of these coordinates of the StarOne C1 satellite state vector is required.\\

\section{Bibliography}

\end{document}